# A novel exact solution of the 2+1-dimensional radial Dirac equation for the generalized Dirac oscillator with the inverse potentials

ZiLong Zhao, ZhengWen Long and MengYao Zhang

Department of Physics, Guizhou University 550025, China

Corresponding author, E-mail: zwlong@gzu.edu.cn

The generalized Dirac oscillator as one of the exact solvable model in quantum mechanics was introduced in 2+1-dimensional world in this paper. What is more, the general expressions of the exact solutions for these models with the inverse cubic, quartic, quintic and sixtic power potentials in radial Dirac equation were further given by means of the Bethe ansatz method. And finally, the corresponding exact solutions in this paper were further discussed.

**PACS**: 03.65.Pm, 03.65.Ge

**Keywords**: radial Dirac equation, generalized Dirac oscillator, Bethe ansatz method, inverse potential, exact solution

## 1. Introduction

As we all know, Dirac oscillator as a concept which was first proposed by M. Moshinsky [1] and collaborators by means of the combination between momentum and coordinate $p \to p - im\omega\beta r$, where $r$ is the coordinates, $m$ represents the mass of particle, and $\omega$ is regarded as the frequency of particle [2-3]. Although the Dirac oscillator, an effective relativistic model, has been succeed in describing the interaction between anomalous magnetic moment and linear coordinates in many fields, it is powerless to explain the interaction between anomalous magnetic moment and nonlinear coordinates. Therefore, in order to describe the complex interactions, Dutta and his colleagues [4] proposed a new concept which is called the generalized Dirac oscillator. So it could generalized the linear effect between coordinates and momentum to nonlinear effect by using this system, that is to say, the Dirac oscillator could be regarded as a special case in the generalized Dirac oscillator.

As one of the few relativistic quantum systems that can be solved accurately, Dirac oscillator has attracted a lot of attentions [1-5]. This system can not only solve the quark confining potential [5-7] in quantum chromodynamics, but also can be used to solve some complex interactions [8-9]. Considering the practicability of solving problems, Dirac oscillator has been extensively discussed by many researchers in various aspects: like conformal invariance properties [8], covariance properties [9], shift operators [10], symmetry Lie algebra [11], hidden supersymmetry [12-14], completeness of wave functions and so on [15-16]. Moreover, many phenomena in Quantum Optics [17-21] could also be explained by this model. The investigation of the relationship between Dirac oscillator and relativistic Jaynes-Cummings model [4, 20-21] bridges the two unrelated fields of Relativistic Quantum Mechanics and Quantum Optics [22-23]. In addition, this model can also be used to explain some new phenomena in condensed matter physics [16-17, 24-25], such as the Quantum Hall Effect and Fractional Statistics. So it is interesting to extend Dirac oscillator to generalized Dirac oscillator for solving the interaction between momentum and coordinate in nonlinear electric field.

To solve 1+1-dimensional Dirac equation with complex interaction potential, Dutta and colleagues constructed the generalized Dirac oscillator by making a simple transformation $p \to p - i\hat{\beta}f(x)$. But now, we want to solve the 2+1-dimensional radial Dirac equation with complex interaction potential, so the momentum operator is transformed $p_x \to p_x - i\hat{\beta}xf(r)$ and $p_y \to p_y - i\hat{\beta}yf(r)$ in this paper. Therefore, the corresponding complete solutions of 2+1-dimensional radial Dirac equation can be given by the Bethe ansatz method [26-29] provided that appropriate selection the interactions.

Next, the structure of this paper is as follows. The 2+1-dimensional generalized Dirac oscillator is introduced in Section 2. In Section 3, exact solutions of the radial Dirac equation with generalized Dirac oscillator are given by using the functional of Bethe ansatz method [26-29] when selecting appropriate interactions. Section 4 is devoted to the conclusion. And finally, Bethe ansatz method will be briefly reviewed in appendix A.

## 2. 2+1-dimensional generalized Dirac oscillator

For particle of mass $M$, the Dirac equation can be written in terms of two component spinors [30-32]

$$E\psi(r,\theta) = \hat{\beta}(\hat{\gamma}^j p_j - M)\psi(r,\theta), \quad j=1,2, \quad (1)$$

where $\hat{\beta}$ and $\hat{\beta}\hat{\gamma}^j$ are matrices that they are constituted by the Pauli spin matrices

$$\hat{\beta}\hat{\gamma}^1 = \hat{\sigma}_1, \quad \hat{\beta}\hat{\gamma}^2 = s\hat{\sigma}_2, \quad \hat{\beta} = \hat{\sigma}_z, \quad (2)$$

and the parameter $s$ takes the values $\pm 1$ (+1 for spin up and $-1$ for spin down). The formalism (1) can transform two uncoupled two-component Dirac equations for $s = +1$ or $s = -1$ by using the decoupling for the four-component Dirac equation in the absence of the third spatial coordinate. Thus, one of the two-component Dirac equations could be used to describe the positive energy eigenstate and another to describe the negative energy eigenstate. And the positive energy eigenstate is what usually called particle state, and the corresponding negative energy eigenstate is called antiparticle state. Due to there is only one spin polarization in 2+1-dimensional space,

so the equation of Dirac oscillator for the particle state can be written as follow [32-33]

$$E\psi(r,\theta) = [\hat{\sigma}^j \cdot (p_j - iM\omega\hat{\beta}\vec{r}) + \hat{\sigma}_z M]\psi(r,\theta), \quad j=1,2 \quad (3)$$

where the matrices and the eigenfunction satisfy the following conditions

$$\psi(r,\theta) = (\phi(r,\theta) \quad \varphi(r,\theta))^T,$$
$$\hat{\sigma}_1 = \hat{\sigma}_x = \begin{pmatrix} 0 & 1 \\ 1 & 0 \end{pmatrix}, \hat{\sigma}_2 = \hat{\sigma}_y = \begin{pmatrix} 0 & -i \\ i & 0 \end{pmatrix}. \quad (4)$$

Next, the momentum operators will be replaced by $p_x \to p_x - i\hat{\beta}xf(r)$ and $p_y \to p_y - i\hat{\beta}yf(r)$ in free radial Dirac equation, so that corresponding radial Dirac equation of generalized oscillator is given as follow

$$E\begin{pmatrix}\phi\\\varphi\end{pmatrix} = [\hat{\sigma}_x(p_x - i\hat{\beta}xf(r)) + \hat{\sigma}_y(p_y - i\hat{\beta}yf(r)) + \hat{\sigma}_z M]\begin{pmatrix}\phi\\\varphi\end{pmatrix}. \quad (5)$$

According to the equation (4), the equation (5) can be written as follow

$$(E-M)\phi = [p_x + ixf(r) - ip_y + yf(r)]\varphi,$$
$$(E+M)\varphi = [p_x - ixf(r) + ip_y + yf(r)]\phi. \quad (6)$$

So, the second order differential equation could be given by substituting the lower formula in equation (6) into the upper formula

$$\left[r\frac{\partial f(r)}{\partial r} - \frac{1}{r}\frac{\partial}{\partial r}\left(r\frac{\partial}{\partial r}\right) - \frac{1}{r^2}\frac{\partial^2}{\partial \theta^2} - \varepsilon^2 - 2L_z f(r) + r^2 f^2(r) + 2f(r)\right]\varphi = 0, \quad (7)$$

where $\varepsilon^2 = (E^2 - M^2)$. Next, the exact solution of this system will be further discussed in following section by using the Bethe ansatz method.

## 3. Explicit implementation of generalized Dirac oscillator

Over the past decades, the singular potential has attracted lots of attentions because it can be used to describe many physical problems. For example, the singular potential is not only widely applied in the research of the $(p, p)$ and $(p, \pi)$ procedures in high-energy physics [29, 34-35], but also repulsive singular potentials could reproduce the interactions of $\alpha - \alpha$ scattering and nucleons with K-mesons [29, 36]. Moreover, high energy scattering question caused by strong singular potential, as an example of non-relativistic quantum mechanics, has also been extensively discussed by many authors [29, 37-38]. Besides, the singular potentials were also widely used to describe the interaction of two atoms in molecular physics, inter-atomic or intermolecular force and chemical physics [29]. Considering the wide applications of the singular potential, the exact solutions of 2+1-dimensional radial Dirac equation with generalized Dirac oscillator under the inverse cubic, quartic, quintic and sixtic power potentials were discussed in this paper. And it is also shown that the conclusion of the generalized Dirac oscillator with higher order inverse potential can degenerate to the conclusion of lower order inverse power potential interaction when the parameters are properly selected. Next, the analytical expressions of the corresponding exact solutions will be given in section 3, the ground and first excited state was discussed in more detail.

## 3.1. Quasi exact solution of the inverse cubic power potential

Now, let's firstly discuss the inverse cubic power potential

$$f(r) = a + \frac{b}{r} + \frac{d}{r^2} + \frac{e}{r^3}. \tag{8}$$

This potential has been investigated in Schrodinger equation by the other authors to obtain the accurate analytical expression [39-40]. Now, in order to solve this problem in the 2+1-dimensional radial Dirac equation, the corresponding radial Dirac equation could be given by substituting equation (8) into equation (7)

$$\left[ -\frac{1}{r}\frac{\partial}{\partial r}\left(r\frac{\partial}{\partial r}\right) - \frac{1}{r^2}\frac{\partial^2}{\partial \theta^2} + a^2 r^2 \right.$$
$$+ 2abr + 2i\left(a + \frac{b}{r} + \frac{d}{r^2} + \frac{e}{r^3}\right)\frac{\partial}{\partial \theta}$$
$$+ 2ad + b^2 + 2a - \varepsilon^2 + \frac{2a(ae+bd)+b}{r} \tag{9}$$
$$\left. + \frac{2be+d^2}{r^2} + \frac{(2d-1)e}{r^3} + \frac{e^2}{r^4} \right]\varphi = 0.$$

The exact solution of the 2+1-dimensional radial Dirac equation is further given by extracting the asymptotic behavior of the wave function $\varphi(r,\theta)$ through a simple replacement. Next, by making a brief examination for differential equation, we implement transformation

$$\varphi(r,\theta) = r^\alpha g(r)\exp\left(Ar^2 + Br + D/r - im\theta\right), \tag{10}$$

here the parameters $\alpha, A, B$ and $D$ are constant and $m$ is magnetic quantum number. Substituting equation (10) into differential equation (9), it's easy to find that parameters satisfy underlying relations

$$\alpha = 1 - m - d > 0, A = a/2, D = e.$$

Therefore, new differential equation could be written as follow

$$r^2 g''(r) + \left[2ar^3 + 2Br^2 + 2\alpha r + r - 2e\right]g'(r)$$
$$+ \left[\left(4a\alpha - 2a + \varepsilon^2 + B^2 - b^2\right)r^2 + (2\alpha B + B \tag{11}\right.$$
$$\left. - 4ae - \lambda_1 b\right)r + 2(\alpha - be - Be) - 1\right]g(r) = 0,$$

here parameters satisfy $\lambda_1 = 1 + 2m + 2d$. Obviously, the equation (11) could also be solved accurately by using the Bethe ansatz method [26-29] if the potential parameters satisfy certain constraints. Now, in order to realize the square integrable of wave functions, we assume that $B = 0$ or $a = 0$, then the degree $n$ polynomials solutions of the equation (11) could be given

$$g(r) = \prod_{j=1}^{n}(r - r_j), g(r) \equiv 1 \text{ for } n = 0. \tag{12}$$

*Case one:* $B = b = 0$

In this section, the equation (11) would be transformed into the following form

$$r^2 g''(r) + \left(2ar^3 + 2\alpha r + r - 2e\right) g'(r) - \left[4aer + 1 - 2\alpha - \left(4a\alpha + \varepsilon^2 - 2a\right) r^2\right] g(r) = 0. \tag{13}$$

Substituting equation (12) into equation (13) and using Bethe ansatz method, it is easy to prove that parameters satisfy the following constraints

$$2a(2\alpha + n - 1) + \varepsilon^2 = 0, \quad 4ae = 2a \sum_{j=1}^{n} r_j,$$

$$2a \sum_{j=1}^{n} r_j^2 + (2\alpha + n - 1)(n + 1) = 0.$$

And Bethe ansatz equation can be given as follow

$$\sum_{j \neq k}^{n} \frac{2}{r_j - r_k} + \frac{2\left(ar_j^3 + \alpha r_j - e\right) + r_j}{r_j^2} = 0, \quad j = 1, 2, \dots, n.$$

Therefore, the analytical expressions of the wave functions and energy spectrum could be written as

$$E_n^2 = M^2 - 2a(2\alpha - 1 + n),$$

$$\varphi_n(r, \theta) = r^\alpha \left[\prod_{j=1}^{n} (r - r_j)\right] \exp\left(\frac{ar^2}{2} + \frac{e}{r} - im\theta\right). \tag{14}$$

The above wave functions are square integrable, so the corresponding normalization constants could also be given through the standard integral [41]

$$\int_0^\infty r^\nu \exp\left(-\mu_1 r^2 - \frac{\mu_2}{r}\right) dr$$

$$= \frac{\mu_2^\nu}{2^{\nu+1} \sqrt{\pi \mu_1}} G_{02}^{10}\left(\frac{1}{2}, \frac{1}{2} - \frac{\nu}{2}, -\frac{\nu}{2} \bigg| \frac{1}{4} \mu_1 \mu_2^2\right),$$

where has $\mathrm{Re}(\nu) > 0, \mathrm{Re}(\mu_1) > 0, \mathrm{Re}(\mu_2) > 0$, and $G_{pq}^{mn}$ is the Meijer G-function. Thus, the total wave functions could be written as

$$\psi_n(r, \theta) = \Gamma r^\alpha \left[\prod_{j=1}^{n} (r - r_j)\right] \exp\left(\frac{ar^2}{2} + \frac{e}{r} - im\theta\right),$$

and here the matrix term of the total wave function was set as $\Gamma = \left(\dfrac{p_x + ixf(r) - ip_y + yf(r)}{(E - M)} \quad 1\right)^T$.

Next, the corresponding solution of ground state and first excited state could be given directly by using the expression of the exact solution. Obviously, the polynomial solution of the equation (13) satisfies $g(r) = 1$ if we take $n = 0$. Then, the analytical solution of the ground state could be given as follow

$$E_0^2 = M^2,$$
$$\psi_0(r, \theta) = \Gamma r^\alpha \exp\left(ar^2/2 + e/r - im\theta\right), \tag{15}$$

with $2\alpha - 1 = 0$, $4ae = 0$. The case of $n = 1$ corresponds to the first excited state solution of this system, and the analytical expressions of the wave function and the energy spectrum are further given

$$E_1^2 = M^2 - 4a\alpha,$$
$$\psi_1(r, \theta) = \Gamma r^\alpha (r - r_1) \exp\left(ar^2/2 + e/r - im\theta\right), \tag{16}$$

in which the potential parameters meet the needs of following constraints

$$4ae = 2ar_1, \quad 2ar_1^2 + 4\alpha = 0.$$

*Case two:* $A = a/2 = 0$

In this section, we assume that $A = a/2 = 0$. Obviously, if the potential parameters satisfy certain constraints, the exact solution of the system could be given by using the Bethe ansatz method [26-29]. And the condition of the wave function has an acceptable asymptotic behavior is that the parameter B must be negative when $r \to \infty$. So this differential equation reads

$$r^2 g''(r) + \left(2Br^2 + 2\alpha r + r - 2e\right) g'(r) + \left[2\alpha + (2\alpha B + B - \lambda_1 b) r - 2e(B + b) - 1\right] g(r) = 0, \tag{17}$$

where $B^2 + \varepsilon^2 - b^2 = 0$. Substituting equation (12) into equation (17), the following relations could be given by using the Bethe ansatz method

$$B = \lambda_1 b / (1 + 2\alpha + 2n),$$

$$(2\alpha + n - 1)(n + 1) + 2B \sum_{j=1}^{n} r_j = 2e(b + B),$$

and the Bethe ansatz equation could be written

$$\sum_{j \ne k}^{n} \frac{2}{r_j - r_k} + \frac{2(Br_j^2 + \alpha r_j - e) + r_j}{r_j^2} = 0, \; j = 1, 2, ..., n.$$

As mentioned above, the parameter $B$ is negative, so the parameter $b$ must also be negative. Therefore, the analytical expressions of the wave functions and energy spectrum could be written as

$$E_n^2 = M^2 + \left[1 - \left(\frac{3-2\alpha}{1+2n+2\alpha}\right)^2\right]b^2,$$

$$\varphi_n(r,\theta) = r^\alpha \left[\prod_{j=1}^{n}(r-r_j)\right]\exp\left(Br + \frac{e}{r} - im\theta\right). \quad (18)$$

The wave functions $\varphi_n(r,\theta)$ are square integrable, and its corresponding normalization constants could be determined by standard integral [41]

$$\int_0^\infty r^\nu \exp(-\mu_1 r - \mu_2/r) dr = 2(\mu_2/\mu_1)^{(\nu+1)/2},$$

$$\text{Bessel} \quad K = \left(\nu+1, 2\sqrt{\mu_1 \mu_2}\right)$$

for $\operatorname{Re}(\nu) > 0, \operatorname{Re}(\mu_1) > 0, \operatorname{Re}(\mu_2) > 0$. So, the total wave functions were written

$$\psi_n(r,\theta) = \Gamma r^\alpha \left[\prod_{j=1}^{n}(r-r_j)\right]\exp\left(Br + \frac{e}{r} - im\theta\right).$$

Next, the ground and the first excited state will be investigated in detail. Obviously, the polynomial solution of the equation (17) could be written as $g(r) = 1$ for the case $n = 0$, then, corresponding energy spectrum and wave function are

$$E_0^2 = M^2 + 8(2\alpha - 1)\left[b/(1+2\alpha)\right]^2,$$

$$\psi_0(r,\theta) = \Gamma r^\alpha \exp(Br + e/r - im\theta),$$

where the constraint satisfy $2\alpha = 1 + 2e(b+B)$.

The case of $n = 1$ corresponds to first excited state solution of this system, the analytical expressions of wave function and energy spectrum are further given

$$E_1^2 = M^2 + 24\alpha\left[b/(3+2\alpha)\right]^2,$$

$$\psi_1(r,\theta) = \Gamma r^\alpha (r-r_1)\exp(Br + e/r - im\theta),$$

with underlying constraint for potential parameters

$$2\alpha + Br_1 = e(B+b).$$

Here the root $r_1$ can be further given by using the Bethe ansatz equation

$$2Br_1^2 + (2\alpha + 1)r_1 - 2e = 0$$

$$\Rightarrow r_1 = \left(-2\alpha - 1 \pm \sqrt{(2\alpha+1)^2 + 16Be}\right)\Big/4B.$$

### 3.2. Quasi exact solution of the inverse quartic power potential

The interaction of the Dirac equation with inverse quartic power potential is studied in this part

$$f(r) = a + \frac{b}{r^2} + \frac{e}{r^4}, \; a < 0, \; e < 0. \quad (19)$$

From the phenomenological point of view, singular potential as a very useful form of anharmonicity is used in many aspects of physics [42-43]. The inverse quartic potential has also been investigated in many different questions by lots of authors [44-46]. Now, the purpose of our study is to obtain the analytical properties of the scattering amplitude about singular potential. So, the corresponding 2+1-dimensional radial Dirac equation with inverse quartic power potential is given by equation (7)

$$\left[-\frac{1}{r}\frac{\partial}{\partial r}\left(r\frac{\partial}{\partial r}\right) + 2i\left(a + \frac{b}{r^2} + \frac{e}{r^4}\right)\frac{\partial}{\partial \theta}\right.$$
$$-\frac{1}{r^2}\frac{\partial^2}{\partial \theta^2} + a^2 r^2 + (2ab + 2a - \varepsilon^2) \quad (20)$$
$$\left.+\frac{2ae + b^2}{r^2} + \frac{2e(b-1)}{r^4} + \frac{e^2}{r^6}\right]\varphi(r,\theta) = 0.$$

In order to deal with the question, the appropriate asymptotic behavior of the wave function $\varphi(r,\theta)$ is extracted through a simple replacement

$$\varphi(r,\theta) = r^{2-m-b} h(r) \exp\left(ar^2/2 + e/2r^2 - im\theta\right),$$

where the parameter satisfy $2 - m - b > 0$. So, this differential equation for $h(r)$ reads

$$h''(r) + \left[2ar^2 + 3 + 2\lambda_2 - (2e/r^2)\right]h'(r)/r$$
$$+ \left[\varepsilon^2 + 4a\lambda_2 + (4\lambda_2 - 4ae)/r^2\right]h(r) = 0, \quad (21)$$

where the parameter is set $\lambda_2 = 1 - m - b$. Here, the equation (21) can be transformed into a solvable form by making a new variable $t = r^2$

$$4t^2 h''(t) + 4\left[at^2 + (2+\lambda_2)t - e\right]h'(t) \\ + \left[(\varepsilon^2 + 4a\lambda_2)t + 4(\lambda_2 - ae)\right]h(t) = 0. \quad (22)$$

So, the degree $n$ polynomial solutions for the differential equation read

$$h(t) = \prod_{j=1}^{n}(t - t_j), \; h(t) \equiv 1 \text{ for } n = 0 \quad (23)$$

with different roots $\{t_j\}$ if potential parameters fulfill certain constraints.

Thus the analytical expressions for energy states and wave functions could be given

$$E_n^2 = M^2 - 4a(n + \lambda_2),$$
$$\varphi_n(r,\theta) = r^{2-m-b}\left[\prod_{j=1}^{n}(r^2 - t_j^2)\right] \quad (24)$$
$$\times \exp(ar^2/2 + e/2r^2 - im\theta).$$

And the parameters satisfy the restrictive conditions

$$a\sum_{j=1}^{n} t_j + (n+1)(n+\lambda_2) = ae,$$

the roots $\{t_j\}$ could be determined by the Bethe ansazt equations

$$\sum_{j \ne k}^{n} \frac{2}{t_j - t_k} + 4\frac{at_j^2 + (\lambda_2 + 2)t_j - e}{t_j^2} = 0, \; j = 1, 2, ..., n.$$

The wave functions are square integrable, and its corresponding normalization constants could be determined by standard integral [41]

$$\int_0^\infty r^\nu \exp\left[-\mu_1 r^2 - (\mu_2/r^2)\right] dr = (\mu_2/\mu_1)^{(\nu+1)/4},$$
$$\text{Bessel} \quad k\left((\nu+1)/2, 2\sqrt{\mu_1 \mu_2}\right)$$

for $\mathrm{Re}(\nu) > 0, \mathrm{Re}(\mu_1) > 0, \mathrm{Re}(\mu_2) > 0$.

As special cases of general expressions, we focus on studying ground state and first excited state systems in this section. First, let's give the corresponding wave function and ground state energy spectrum in terms of the equation (24)

$$E_0^2 = M^2 - 4a\lambda_2,$$
$$\psi_0(r,\theta) = \Gamma r^{2-m-b}\exp(ar^2/2 + e/2r^2 - im\theta), \quad (25)$$

where the potential parameters comply with $\lambda_2 = ae$.

The case of $n = 1$ corresponds to first excited state solution of the system, and the analytical expressions of the wave functions and energy spectrum for the first excited state are further written

$$E_1^2 = M^2 - 4a(1 + \lambda_2),$$
$$\psi_1(r,\theta) = \Gamma r^{2-m-b}(r^2 - t_1^2) \quad (26)$$
$$\times \exp(ar^2/2 + e/2r^2 - im\theta),$$

in which the different roots $\{t_j\}$ could be computed analytically by the Bethe ansazt equation [26-29]

$$at_1^2 + (2+\lambda_2)t_1 - e = 0$$
$$\Rightarrow t_1 = \left(m + b - 3 \pm \sqrt{(2+\lambda_2)^2 + 4ae}\right)\Big/2a,$$

and the potential parameters satisfy the following restrictive condition

$$(3m + 3b + 2ae - 5)^2 = (3 - m - b)^2 + 4ae.$$

### 3.3. Quasi exact solution of the inverse quintic power potential

Now, the inverse quintic power potential will be discussed in there

$$f(r) = a + \frac{b}{r} + \frac{d}{r^2} + \frac{e}{r^3} + \frac{f}{r^4} + \frac{g}{r^5}, \; g < 0. \quad (27)$$

The exact solvable form of the radial Dirac equation with inverse quintic power potential could be further given by substituting equation (27) into equation (7)

$$\left[-\frac{1}{r}\frac{\partial}{\partial r}\left(r\frac{\partial}{\partial r}\right)-\frac{1}{r^2}\frac{\partial^2}{\partial\theta^2}+2if(r)\frac{\partial}{\partial\theta}\right.$$
$$+a^2r^2+2a(br+1+\lambda_3)+b^2-\varepsilon^2$$
$$+\frac{b+2b\lambda_3+2ae}{r}+\frac{\lambda_3^2+2af+2be}{r^2}$$
$$+\frac{2(e\lambda_3+ag+bf)-e}{r^3}+\frac{2eg+f^2}{r^6} \quad (28)$$
$$+\frac{2(f\lambda_3-f+bg)+e^2}{r^4}+\frac{2fg}{r^7}$$
$$\left.+\frac{(2\lambda_3-3)g+2ef}{r^5}+\frac{g^2}{r^8}\right]\varphi=0,$$

where the parameter satisfy $\lambda_3 = m+d$. Similar to the studies in preceding sections, the corresponding exact solutions can be given provided that the appropriate asymptotic behavior of the wave function is extracted. Here we set

$$\varphi(r,\theta) = r^\eta k(r)\exp(Ar^2+Br \\ +D/r+F/r^2+G/r^3-im\theta), \quad (29)$$

where the parameters $\eta, A, B, D, F, G$ are constant and parameter $m$ is the magnetic quantum number. Substituting equation (29) into equation (28), we find that the parameters satisfy

$$\eta = 3-m-d > 0, \quad A = a/2,$$
$$D = e, \quad F = f/2, \quad G = g/3.$$

Thus, the radial differential equation (28) will degenerate into the following form

$$r^4 k''(r)+\left[2(ar^5+Br^4-er^2-fr-g)\right.$$
$$+(7-2\lambda_3)r^3\right]k'(r)+\left\{\left[B^2+6a+\varepsilon^2-b^2\right.\right.$$
$$-4a\lambda_3\right]r^4-\left[2(B\lambda_3+2ae+b\lambda_3)+b \quad (30)$$
$$-7B\right]r^3+(9-4af-6\lambda_3-2e\lambda_4)r^2-(4e$$
$$+4ag+2f\lambda_4)r-(2f+2g\lambda_4)\right\}k(r)=0,$$

where $\lambda_4 = B+b$. It's easy to find that equation (30) is exactly solvable when making $B = b = 0$ or $a = 0$, and its corresponding exact solutions could also be given by the degree $n$ polynomials

$$k(r) = \prod_{j=1}^n (r-r_j), k(r) \equiv 1 \text{ for } n=0. \quad (31)$$

*Case one*: $B = b = 0$

Considering this case of $B = b = 0$, the radial differential equation (30) can be transformed into

$$r^4 k''(r)+\left[2(ar^5-\lambda_3 r^3-er^2-fr-g)+7r^3\right]$$
$$\times k'(r)+\{(6a-4a\lambda_3+\varepsilon^2)r^4-4aer^3-2f \quad (32)$$
$$+(9-6\lambda_3-4af)r^2-4(e+ag)r\}k(r)=0.$$

Substituting equation (31) into equation (32) and using the Bethe ansatz method, the requirements of the parameters as follow

$$2a(n+3-2\lambda_3)+\varepsilon^2 = 0, \quad 2a\sum_{j=1}^n r_j = 4ae,$$

$$2a\sum_{j=1}^n r_j^2 + (n+3)(n+3-2\lambda_3) = 4af,$$

$$2a\sum_{j=1}^n r_j^3 + (5+2n-2\lambda_3)\sum_{j=1}^n r_j = 2e(n+2)+4ag,$$

$$2a\sum_{j=1}^n r_j^4 + (5+2n-2\lambda_3)\sum_{j=1}^n r_j^2$$
$$+2\sum_{j<k}^n r_j r_k = 2e\sum_{j=1}^n r_j + 2f(1+n).$$

Here the roots $\{r_j\}$ fulfill Bethe ansatz equation

$$\sum_{j\neq k}^n \frac{2}{r_j-r_k}+\frac{2(ar_j^5-\lambda_3 r_j^3-er_j^2-fr_j-g)+7r_j^3}{r_j^4} = 0,$$

where $j = 1, 2, ..., n$. And its corresponding exact solutions could be written as following

$$E_n^2 = M^2 - 2a(3+n-2\lambda_3),$$
$$\psi_n(r,\theta) = \Gamma r^\eta \left[\prod_{j=1}^n (r-r_j)\right]\exp(ar^2/2 \quad (33)$$
$$+e/r+f/2r^2+g/3r^3-im\theta).$$

The wave functions are also squarely integrable: $\int_0^\infty |\psi_n(r)|^2 dr < \infty$. Next, the ground and first excited state will be discussed in detail as two special cases. And the case of $n=0$ corresponds to the ground state solution of the system

$$E_0^2 = M^2,$$
$$\psi_0(r,\theta) = \Gamma r^\eta \exp\left(ar^2/2 + e/r \right. \tag{34}$$
$$\left. + f/2r^2 + g/3r^3 - im\theta\right).$$

The first excited state solution can be given by making $n=1$

$$E_1^2 = M^2 - 4a(2-\lambda_3),$$
$$\psi_1(r,\theta) = \Gamma r^\eta (r-r_1)\exp\left(ar^2/2 + e/r \right. \tag{35}$$
$$\left. + f/2r^2 + g/3r^3 - im\theta\right),$$

in which the root $r_1$ could be computed analytically by Bethe ansatz equation

$$2ar_1^5 + (7-2\lambda_3)r_1^3 - 2er_1^2 - 2fr_1 - 2g = 0,$$

and its parameters satisfy the following relations

$$2ar_1 = 4ae, \quad ar_1^2 + 8 = 2af + 4\lambda_3,$$
$$2ar_1^3 + (7-2\lambda_3)r_1 = 6e + 4ag,$$
$$2ar_1^4 + (7-2\lambda_3)r_1^2 = 2er_1 + 4f.$$

*Case two:* $A = a/2 = 0$

Here, we set $A = a/2 = 0$. In order to get acceptable asymptotic behavior for the wave function when $r \to \infty$, the parameter $B$ is required to be negative. Furthermore, according to the Bethe ansatz method, we require that $\varepsilon^2 + B^2 - b^2 = 0$. Thus, the differential equation (30) could be written as

$$r^4 k''(r) + \left[2\left(Br^4 - \lambda_3 r^3 - er^2 - fr - g\right)\right.$$
$$\left. + 7r^3\right]k'(r) + \left\{[7B - 2(B\lambda_3 + b\lambda_3)\right.$$
$$\left. - b]r^3 + (9 - 6\lambda_3 - 2e\lambda_4)r^2 - 2(2e \right. \tag{36}$$
$$\left. + f\lambda_4)r - 2(f + g\lambda_4)\right\}k(r) = 0.$$

Substituting equation (31) into equation (36) and using the Bethe ansatz method, the exact solutions of equation (36) can be given as following

$$E_n^2 = M^2 + \frac{(6+2n-4\lambda_3)^2 b^2}{(7+2n-2\lambda_3)^2},$$
$$\psi_n(r,\theta) = \Gamma r^\eta \left[\prod_{j=1}^{n}(r-r_j)\right]\exp\left(Br \right. \tag{37}$$
$$\left. + e/r + f/2r^2 + g/3r^3 - im\theta\right),$$

and the parameters satisfy the determined constraints

$$(7+2n-2\lambda_3)B = b(1+2\lambda_3),$$
$$2B\sum_{j=1}^{n} r_j + (n+3)(3+n-2\lambda_3) = 2e\lambda_4,$$
$$2B\sum_{j=1}^{n} r_j^2 + (5+2n-2\lambda_3)\sum_{j=1}^{n} r_j = 2(ne + 2e + \lambda_4 f),$$
$$2B\sum_{j=1}^{n} r_j^3 + (5+2n-2\lambda_3)\sum_{j=1}^{n} r_j^2 + 2\sum_{j<k}^{n} r_j r_k$$
$$= 2(nf + f + g\lambda_4) + e\sum_{j=1}^{n} r_j.$$

Here the roots satisfy Bethe ansatz equation

$$\sum_{j \neq k}^{n} \frac{2}{r_j - r_k} + \frac{2\left(Br_j^4 - \lambda_3 r_j^3 - er_j^2 - fr_j - g\right) + 7r_j^3}{r_j^4} = 0,$$

where $j = 1, 2, ..., n$. As we mentioned above, the parameter $B$ is negative, so the parameter $b$ must be negative. Next, in order to satisfy the wave functions square integrable we assume that $g < 0$. It isn't difficult to use Maple for checking numerically.

Similar to the previous cases, the ground state solution corresponds to the case of $n=0$

$$E_0^2 = M^2 + \left[(6-4\lambda_3)b/(7-2\lambda_3)\right]^2,$$
$$\psi_0(r,\theta) = \Gamma r^\eta \exp\left(Br + e/r \right. \tag{38}$$
$$\left. + f/2r^2 + g/3r^3 - im\theta\right),$$

with the constraints

$$2e + \lambda_4 f = 0, \quad 3(3-2\lambda_3) = 2e\lambda_4,$$
$$f + g\lambda_4 = 0, \quad B = (1+2\lambda_3)b/(7-2\lambda_3).$$

And the first excited state solution corresponds to the case of $n=1$

$$E_1^2 = M^2 + \left[(8-4\lambda_3)b/(9-2\lambda_3)\right]^2,$$

$$\psi_1(r,\theta) = \Gamma r^\eta (r-r_1)\exp(Br \quad (39)$$
$$+ e/r + f/2r^2 + g/3r^3 - im\theta),$$

comply with the constrains

$$Br_1 + 8 = 4\lambda_3 + e\lambda_4, \quad B = (1+2\lambda_3)b/(9-2\lambda_3),$$
$$2Br_1^2 + (7-2\lambda_3)r_1 = 6e + 2\lambda_4 f,$$
$$2Br_1^3 + (7-2\lambda_3)r_1^2 = 6f + 2g\lambda_4 + er_1.$$

Here the root $r_1$ satisfying

$$2Br_1^4 + (7-2\lambda_3)r_1^3 - 2(er_1^2 + fr_1 + g) = 0.$$

### 3.4. Quasi exact solution of the inverse sixtic power potential

The effect of the inverse sixth power potential in radial Dirac equation is discussed as the last case

$$f(r) = a + \frac{d}{r^2} + \frac{f}{r^4} + \frac{h}{r^6}, \quad h < 0. \quad (40)$$

The inverse power potential has played an important role in the study for particles interactions in atomic, molecular and nuclear physics [47-49]. Substituting equation (40) into equation (7), the corresponding radial Dirac equation with the inverse sixth power potential could be given

$$\left[-\frac{1}{r}\frac{\partial}{\partial r}\left(r\frac{\partial}{\partial r}\right) + 2i\left(a + \frac{d}{r^2} + \frac{f}{r^4} + \frac{h}{r^6}\right)\frac{\partial}{\partial \theta}\right.$$
$$-\frac{1}{r^2}\frac{\partial^2}{\partial \theta^2} + a^2 r^2 + 2ad + 2a - \varepsilon^2$$
$$+ \frac{2af + d^2}{r^2} + \frac{2(ah+df-f)}{r^4} + \frac{2fh}{r^8} \quad (41)$$
$$\left. + \frac{h^2}{r^{10}} + \frac{2(dh-2h)+f^2}{r^6}\right]\varphi(r,\theta) = 0.$$

Similar to the discussions in previous sections, in order to extract appropriate asymptotic behavior from the wave function $\varphi(r,\theta)$, we setting

$$\varphi(r,\theta) = r^\beta y(r)\exp\left(\frac{ar^2}{2} + \frac{f}{2r^2} + \frac{h}{4r^4} - im\theta\right), \quad (42)$$

where the parameter satisfy $\beta = 4 - m - d$. Then, according to equation (42), the differential equation for $y(r)$ is written as follow

$$y''(r) + \left(2ar + \frac{2\beta+1}{r} - \frac{2f}{r^3} - \frac{2h}{r^5}\right)y'(r)$$
$$+ \left[2a(\beta - m - d) + \varepsilon^2 - \frac{4(f+ah)}{r^4}\right. \quad (43)$$
$$\left. + \frac{\beta^2 - 4af - (m+d)^2}{r^2}\right]y(r) = 0.$$

Replace the variable with $z = r^2$ and multiply $z^2$ on both sides of the equation, so that the equation can degenerate into the following form

$$4z^3 y''(z) + 4(az^3 + \beta z^2 + z^2 - fz - h)y'(z)$$
$$- \left[4(f+ah) - 4(4 - 2m - 2d - af)z\right. \quad (44)$$
$$\left. - (8a - 4am - 4ad + \varepsilon^2)z^2\right] y(z) = 0.$$

It is easy to find that the degree $n$ polynomial solutions of the differential equation (44) satisfy

$$y(z) = \prod_{j=1}^{n}(z - z_j), \quad y(z) = 1 \text{ for } n = 0. \quad (45)$$

Substituting equation (45) into equation (44) and applying the Bethe ansatz method, the analytical expressions of the energy spectrum and the wave functions could be given

$$E_n^2 = M^2 - 4a(n + 2 - m - d),$$

$$\psi_n(r,\theta) = \Gamma r^\beta \left[\prod_{j=1}^{n}(r^2 - z_j)\right]\exp(ar^2/2 \quad (46)$$
$$+ f/2r^2 + h/4r^4 - im\theta).$$

Potential parameters satisfy the following constraints

$$a\sum_{j=1}^{n} z_j + (n+2)(n+2-m-d) = af,$$

$$a\sum_{j=1}^{n} z_j^2 + (2n+3-m-d)\sum_{j=1}^{n} z_j = (n+1)f + ah.$$

Here the roots $\{z_j\}$ could be given by Bethe ansatz equations

$$\sum_{j \neq k}^{n} \frac{2}{z_j - z_k} + \frac{az_j^3 + (\beta+1)z_j^2 - fz_j - h}{z_j^3} = 0,$$

where $j = 1, 2, ..., n$. The above functions $\psi_n(r)$ are also squarely integrable, i.e. $\int_0^\infty |\psi_n(r)|^2 dr < \infty$. It is not difficult to find that using Maple for numerical checking is easy to implement.

Obviously, the ground state solution could be given by setting $n = 0$

$$E_0^2 = M^2 - 4a(2 - m - d),$$

$$\psi_0(r, \theta) = \Gamma r^\beta \exp\left(\frac{ar^2}{2} + \frac{f}{2r^2} + \frac{h}{4r^4} - im\theta\right), \quad (47)$$

with the constraints

$$2(2 - m - d) = af, \quad f + ah = 0.$$

The analytical expressions of the first excited state energy spectrum and the wave function for the case of $n = 1$ could be given as following

$$E_1^2 = M^2 - 4a(3 - m - d),$$

$$\psi_1(r, \theta) = \Gamma r^\beta \left[(r^2 - z_1)\right] \exp\left(ar^2/2 + f/2r^2 + h/4r^4 - im\theta\right), \quad (48)$$

and the potential parameters satisfy the constrains

$$az_1 + 3(\beta - 1) = af, \quad az_1^2 + (\beta + 1)z_1 = 2f + ah.$$

Here the root $z_1$ can be further given by using the Bethe ansatz equation

$$az_j^3 + (\beta + 1)z_j^2 - fz_j - h = 0.$$

## 4. Conclusion

In the present work, the 2+1-dimensional radial Dirac equation of generalized oscillator is introduced and its corresponding energy spectrum and the wave function are further given by Bethe ansatz method. The cubic, quartic, quintic and sixth inverse power potentials are studied as examples of singular potentials and its general expressions of exact solutions are given. The corresponding energy spectrum and wave function of the ground state and first excited state are further given by using the general expressions of exact solutions of the Dirac equation with generalized Dirac oscillators. It is easy to find that the results in section 3.3 could degenerate into the conclusions for the inverse cubic power potential in section 3.1 when $f, g \to 0$. And the parameter satisfies $h \to 0$, the results in section 3.2 could also be reproduced by the solutions of the inverse sixtic power potential in section 3.4.

## 5. Acknowledgements

This work was supported by the National Nature Science Foundation of China (Grant Nos., 11465006).

## Appendix A. Review of the functional Bethe ansatz method

The exact solutions are given by the functional Bethe ansatz method [26-29] provided that the general second order linear ordinary differential equation could be written as following

$$\left[P(r)\frac{d^2}{dr^2} + Q(r)\frac{d}{dr} + W(r)\right] S(r) = 0, \quad (A.1)$$

where $P(r), Q(r)$ and $W(r)$ are the polynomials of at most degree 4, 5 and 4 respectively, where the variable $r$ is coordinate and the polynomials satisfy certain constraints $\deg Q(r) > \deg W(r)$. The polynomials $P(r), Q(r)$ and $W(r)$ are given in terms of the forms as follow

$$P(r) = \sum_{k=0}^{n} p_k r^k, \quad Q(r) = \sum_{k=0}^{n} q_k r^k, \quad W(r) = \sum_{k=0}^{n} w_k r^k, \quad (A.2)$$

where $p_k, q_k$ and $w_k$ are constants and $n$ is non-negative integers. Due to the formalism of equation (A.1) is exactly solvable, so the exact solutions of the differential equation with respect to octic potentials [50] had been given because whose formalism is analogous to the equation (A.1). To solve the 2+1-dimensional Dirac equation with suitable interactions, we seek polynomial solutions by using the functional Bethe ansatz method [26-29]

$$S(r) = \prod_{j=1}^{n}(r - r_j), \quad S(r) = 1 \text{ for } n = 0, \quad (A.3)$$

here the distinct roots $r_1, r_2, ..., r_n$ can be determined.

And we can easily find the parameters of the equation (A.1) must satisfy the condition

$$-w_0 = \left(p_4 r^4 + p_3 r^3 + p_2 r^2 + p_1 r + p_0\right) \sum_{j=1}^{n} \frac{1}{r-r_j} \sum_{j \neq k}^{n} \frac{2}{r_j - r_k} + \left(q_5 r^5 + q_4 r^4 + q_3 r^3 + q_2 r^2 + q_1 r + q_0\right) \sum_{j=1}^{n} \frac{1}{r-r_j} + w_4 r^4 + w_3 r^3 + w_2 r^2 + w_1 r. \quad (A.4)$$

Next, we can get the following relationship about the constraints of the distinct roots. Because the equation (A.4) on the left is a constant, but the right of this equation is a meromorphic function with simple poles $r = r_j$ and singularity at $r = \infty$, so the condition of the equation established is that the right side of the equation must be a constant. Thus, we ask for the coefficients of the powers of $r$ as well as the residues at the simple poles of the right side are zero. According to Liouville's theorem, the necessary and sufficient condition is that the right side of equation must be a constant. The constraints can be written as follow

$$w_4 = -nq_5, \quad w_3 = -q_5 \sum_{j=1}^{n} r_j - nq_4,$$

$$w_2 = -q_5 \sum_{j=1}^{n} r_j^2 - q_4 \sum_{j=1}^{n} r_j - n(n-1)p_4 - nq_3,$$

$$w_1 = -q_5 \sum_{j=1}^{n} r_j^3 - q_4 \sum_{j=1}^{n} r_j^2 - n(n-1)p_3$$

$$-nq_2 - \left[2(n-1)p_4 + q_3\right] \sum_{j=1}^{n} r_j, \quad (A.5)$$

$$w_0 = -q_5 \sum_{j=1}^{n} r_j^4 - q_4 \sum_{j=1}^{n} r_j^3 - 2p_4 \sum_{j<k}^{n} r_j r_k$$

$$-\left[q_3 + 2(n-1)p_4\right] \sum_{j=1}^{n} r_j^2 - n(n-1)p_2$$

$$-\left[2(n-1)p_3 + q_2\right] \sum_{j=1}^{n} r_j - nq_1.$$

where the roots $\{r_j\}$ satisfy the Bethe ansatz equation

$$\frac{q_5 r_j^5 + q_4 r_j^4 + q_3 r_j^3 + q_2 r_j^2 + q_1 r_j + q_0}{p_4 r_j^4 + p_3 r_j^3 + p_2 r_j^2 + p_1 r_j + p_0} + \sum_{j=1}^{n} \frac{2}{r_j - r_k} = 0, \quad j = 1, 2, ..., n. \quad (A.6)$$